\documentclass[a4paper]{spie}  
\usepackage{graphicx}
\usepackage{cite}

\title{Phase Closure Nulling: results from the 2009 campaign} 

\author{%
Gilles Duvert\supit{a}, 
Fabien Malbet\supit{a}, 
Alain Chelli\supit{a}, 
Rafael Millan-Gabet\supit{b}, \\
John D. Monnier\supit{c}, 
Gail H. Schaefer\supit{d} 
\skiplinehalf
\supit{a} Lab.\ d'Astrophysique de Grenoble (LAOG), UMR 5571 Univ.\
J.\ Fourier/CNRS, BP 53, F-38051 Grenoble cedex 9, France;\\ 
\supit{b} California Institute of Technology, NExScI, MC\,100-22, Pasadena CA
91125, USA;\\ 
\supit{c} Univ.\ of Michigan, 941 Dennison Building, 500 Church
Street, Ann Arbor, MI 48109, USA;\\  
\supit{d} CHARA Array of Georgia State University, Mount Wilson
Observatory, Mount Wilson, CA 91023, USA. } 
\authorinfo{Further author information:\\
  G.D.: E-mail: gilles.duvert@obs.ujf-grenoble.fr, Telephone: +33 476
  514 588 }

\begin{document} 
\maketitle 

\begin{abstract}
  We present here a new observational technique, Phase Closure Nulling
  (PCN), which has the potential to obtain very high contrast
  detection and spectroscopy of faint companions to bright stars. PCN
  consists in measuring closure phases of fully resolved objects with
  a baseline triplet where one of the baselines crosses a null of the
  object visibility function.  For scenes dominated by the
  presence of a stellar disk, the correlated flux of the star around
  nulls is essentially canceled out, and in these regions the
  signature of fainter, unresolved, scene object(s) dominates the
  imaginary part of the visibility in particular the closure
  phase. We present here the basics of the PCN method, the initial
  proof-of-concept observation, the envisioned science cases and
  report about the first observing campaign made on VLTI/AMBER and
  CHARA/MIRC using this technique.
\end{abstract}



\section{Introduction}
\label{sec:intro}  

Phase Closure Nulling (hereafter PCN) was introduced in a paper by
Chelli et al.\ (2009)\cite{2009A&A...498..321C}. In that reference,
the authors did not emphasize the very peculiar series of coincidences
that lead to the technique described here, which in retrospect merit
to be told.
 
\begin{figure}[t]
  \centering
  \includegraphics[width=0.8\columnwidth]{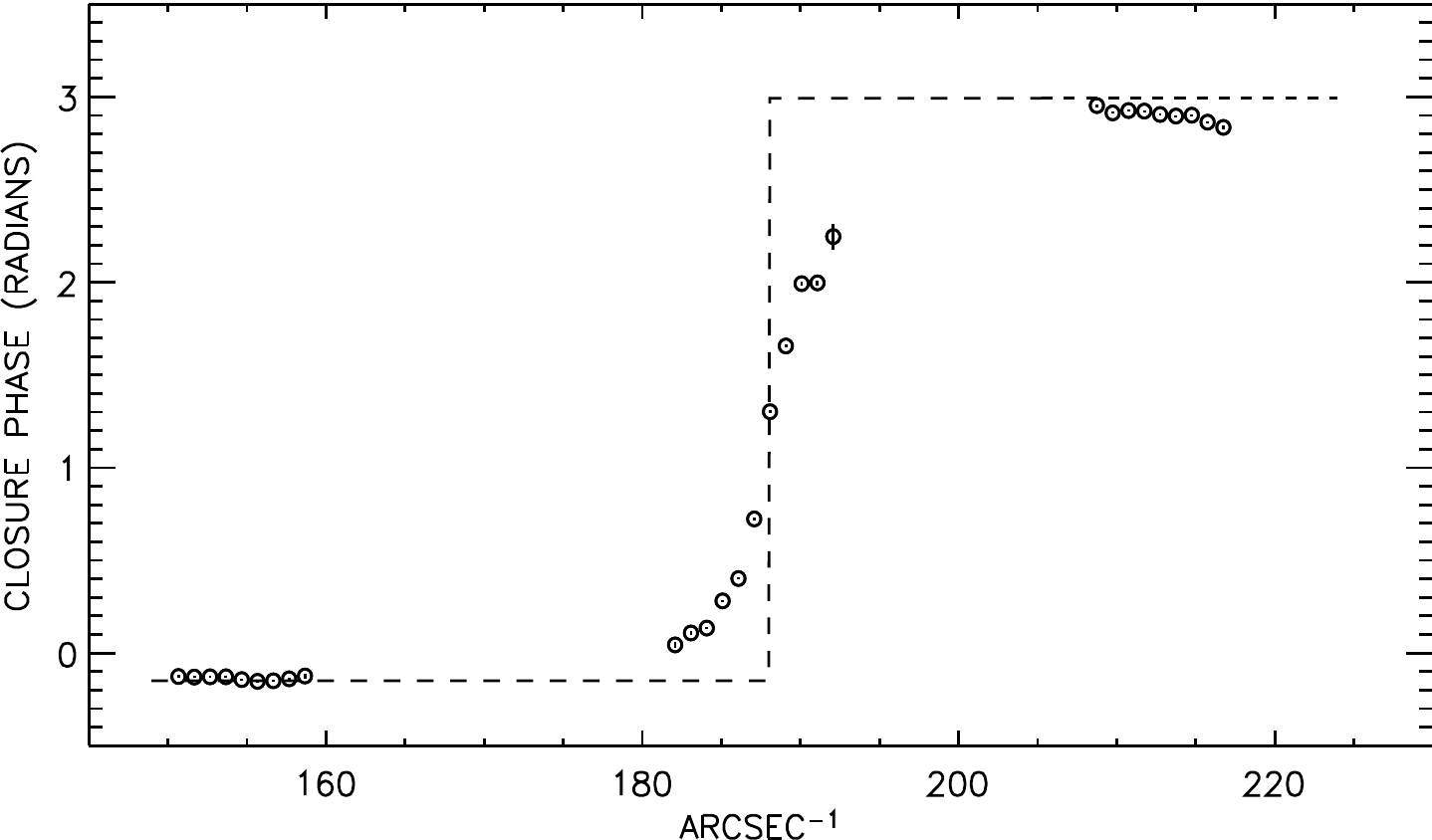}
  \caption{The uncalibrated closure phase of $\sigma$\,Puppis (in
    radians) as a function of spatial frequency. Each of three groups
    of points corresponds to one five minute observation with
    VLTI/AMBER at $R=1500$ in $K$. Thanks to the spectral bandwidth
    used, each observation spans a range of spatial frequencies, so
    the transition from $0$ to $\pi$ radians, as the baseline crosses
    the first null of the stellar disk Bessel function, is
    relatively well sampled. The data was binned to
    $1\,\mathrm{arcsec}^{-1}$ and the error bars, smaller than the
    symbols, are plotted. The overall shape is very different from
    the Heaviside jump expected from a stellar disk, with a smooth
    transition from 0 to $\pi$. Explaining this transition led us to
    develop the theory of Phase Closure Nulling. }
  \label{figure1}
\end{figure}
It really all stems from a few minutes of observations taken in
February 2008 during a night of an AMBER Task Force commissioning run
of \cite{VLT-TRE-AMB-15830-7120}. The purpose of this commissioning
run was to find, and if possible to cure, some obvious instabilities
in the transfer function of AMBER that rendered the absolute
calibration of the visibilities very difficult. We found in particular
that the entrance polarizers (one per beam) were the cause of a
variable fringe beating which biased the visibilities. We thus removed
the polarizers to test on a series of calibrators that the stability
of the instrument was indeed improved. We chose a series of
calibrators from the ESO calibrator list, located at various positions
in the sky and also covering a large range of fluxes. Among these
calibrators, the brightest was the K5III star $\sigma$\,Puppis.

This night run was successful, and demonstrated that the stability of
the instrument was greatly improved without the polarizers. Then, the
polarizers were put back, since otherwise the instrument would not
have been, in ESO terms, nominal. Finally, the culprit polarizers were
replaced by the consortium in Oct, 2008, leading to improved
stability, as expected. Thus, the observations done during our
commissioning run were the one and only ever performed on AMBER
\emph{without polarizers} in the Medium-resolution spectral mode at
$R=1500$.

Of the calibrators used, $\sigma$\,Puppis stood out nicely, because
the closure phase on this star varied from $0$ to $180^{\circ}$ in a
few hours: $\sigma$\,Puppis, being a bright calibrator was also a
large angular-size star and was completely resolved by our $\sim90$\,m
longest baseline, as shown in Fig.~\ref{figure1}. However, the shallow
slope of the closure phase transition from $0$ to $\pi$ radians as the
baseline crosses the first null needed an explanation. Part of the
explanation was quickly found in the literature: $\sigma$\,Puppis was
since 1907 known as a spectroscopic binary and should never have been
included in a calibrator list. Thus, a probable reason for the
deviation from a uniform stellar disk was the presence of the
companion. However, $\sigma$\,Puppis being of the SB1 type, the
companion must be quite faint. The hypothesis that faint companions
could have a visible imprint on the closure phase near the null led us
to the development of the phase closure nulling concept.
\section{Phase Closure Nulling: Theory}

An important property of stars resolved by long baseline
interferometry is that the coherence of the light decreases with
increasing spatial frequencies down to zero before increasing again
following the well known behavior of the Bessel functions.  Michelson
and Pease\cite{1921ApJ....53..249M} used this property to measure the
diameter of Betelgeuse for the first time.

\subsection{The uniform disk case}

For a star represented by an uniform disk of radius $R_\star$, the
visibility is proportional to a Bessel function of the first order,
that is
\begin{equation}
  \label{eq:UD}
  V_\star(u) = 2\,\frac{J_1 (2\pi\,u\,R_\star)}{2\pi\,u\,R_\star}.
\end{equation}
\begin{figure}[t]
  \centering
  \includegraphics[width=0.8\columnwidth]{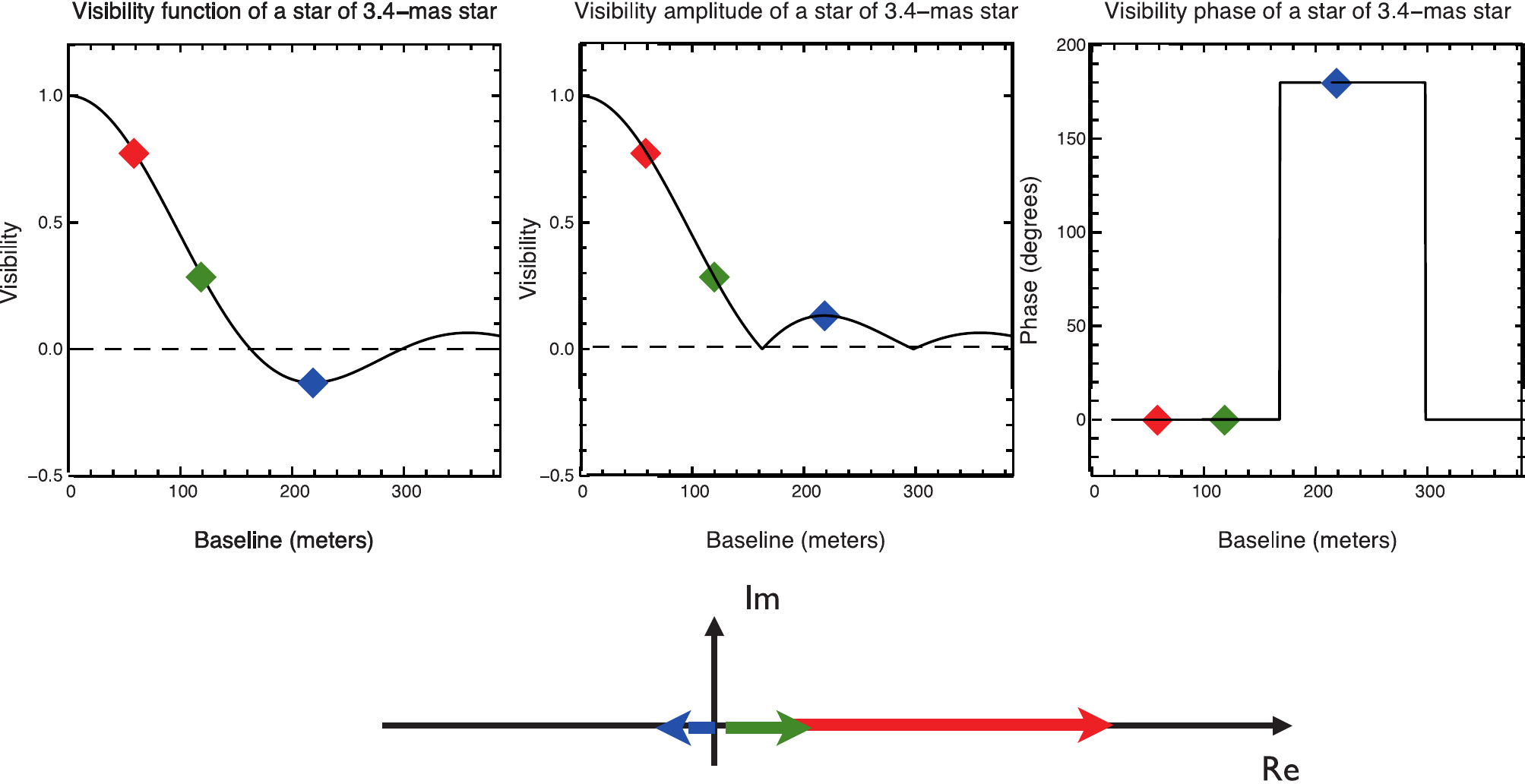}
  \caption{Sketch of the visibity function for a single star, showing
    the origin of the $\pi$ phase shift when one crosses a
    null. Left: complex visibility; middle: visibility amplitude;
    right: phase. Three points, selected at short, middle and high
    spatial frequencies, are represented as colored lozenges. The
    three corresponding complex visibilities are pictured at the
    bottom as Fresnel vectors in an Argand diagram showing the
    complex plane. }
  \label{PCN_singlestar}
\end{figure}
As shown in Fig.~\ref{PCN_singlestar}, $V_\star(u)$ is zero and the
correlated flux of the star is canceled out at spatial frequencies
multiples of $0.61/R_\star$.  Another property is that the phase of
the visibility jumps by 180 degrees at each crossing of the Bessel
function zeros (hereafter called visibility nulls or simply nulls).

\subsection{Perturbation by a close faint companion}

If one adds a faint unresolved companion to the
previous star, assuming for simplicity that the system orientation is
parallel to the frequency axis, the visibility becomes
\begin{equation}
  \label{eq:binary}
  \hat{\mbox{\em \i}}(u)=\frac{V_\star(u)+r e^{i\,2\pi\,us}}{1+r} \,,
\end{equation}
where $r$ is the flux ratio and $s$ is the separation between the two
components. For small flux ratios $r$, the amplitude of the visibility
is only slightly modified by the presence of the companion. The
stronger effect occurs around visibility nulls of the primary where
the visibility perturbation is of the order of $r$. This effect
remains weak and is beyond the performances of current interferometers
as soon as the flux ratio is smaller than 1\%.

More interesting is the phase of the object visibility. The tangent of
the phase of the previous system is given by
\begin{equation}  
\tan\phi(u) = \frac{r\,\sin(2\pi\,us)}{V_\star(u) +
    r\,\cos(2\pi\,us)}.
\label{eq:phase}
\end{equation}
\begin{figure}[t]
  \centering
  \includegraphics[width=0.8\columnwidth]{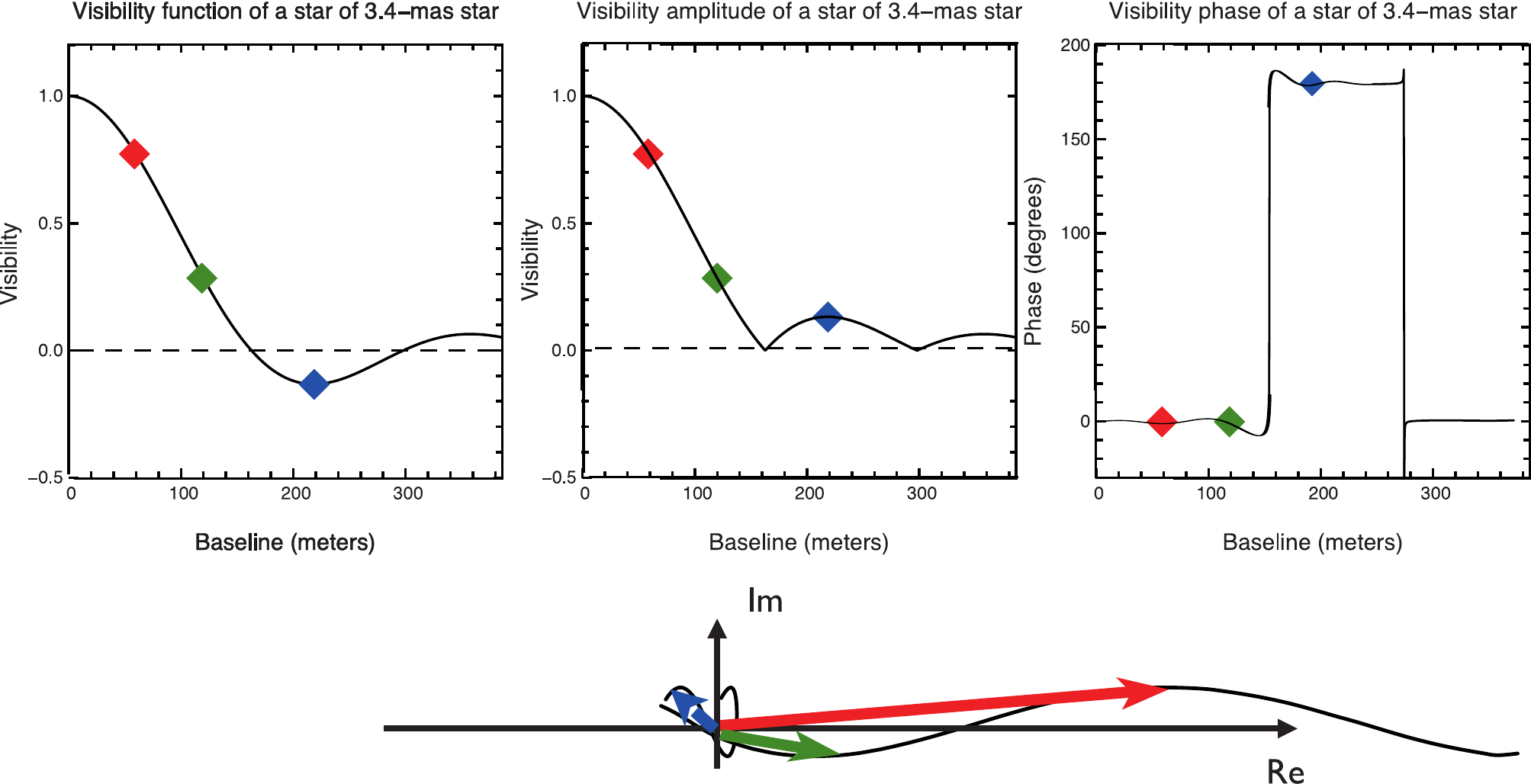}
  \caption{Sketch of the visibility function for a star plus faint
    companion, with the same layout as Fig.~\ref{PCN_singlestar}. The
    bottom Argand diagram displays the corresponding Fresnel
    vectors. The imaginary part due to the faint companion induces a
    phase of the visibility that can become quite large near the
    visibility null at the origin.}
  \label{PCN_doublestar}
\end{figure}
As exemplified in Fig.~\ref{PCN_doublestar}, the phase is the result
of two contributions: from the star $V_\star(u)$ and from the
companion $re^{2i\pi us}$. Except around visibility nulls of the
primary, that is most of the time, the companion produces a phase
signature in the range $\pm r/V_\star(u)$, which remains small.
However, around visibility nulls, in the frequency ranges for which
$|V_\star (u)|\le r$, the phase signature of the companion becomes
significant, with an exact value of $2\pi u_0s$ at the frequencies
$u_0$ of the nulls ($\approx \pi s/R_\star$ for the first null), much
greater than 180 degrees. It follows that, as opposed to the
visibility amplitude, even for small flux ratios, there is always a
frequency interval around nulls within which the phase signature of
the companion is larger that any systematic error and is thus
measurable.

\subsection{The phase closure properties}

Unfortunately, the absolute phase of an interferogram is a quantity
difficult to measure as it requires an absolute reference that in
general does not exist. But, with 3 or more telescopes, one can use
instead the closure phase $\phi_c$ defined as the phase of the
bispectrum on 3 baselines $\hat{I}(u_{12},u_{23})$, with
\begin{equation}
  \hat{I}(u_{12},u_{23})=<\hat{\mbox{\em \i}}(u_{12})\hat{\mbox{\em
      \i}}(u_{23}){\hat{\mbox{\em \i}}}^*(u_{13})> 
  \, ,
\label{bispectrum}
\end{equation}     
where $u_{12}$, $u_{23}$ and $u_{13}$ are the 3 frequencies
transmitted by the interferometer with $u_{13}=u_{12}+u_{23}$, $^*$
denotes the complex conjugate, and $<>$ represents an ensemble
average. The closure phase is a self calibrated observable that unlike
the phase does not need an absolute reference. In addition, it
coincides to the closure phase of the observed object, that is
\begin{equation}
\phi_c=\phi_o(u_{12})+\phi_o(u_{23})-\phi_o(u_{13}) \, ,
\end{equation}     
where $\phi_o$ is the phase of the object spatial Fourier transform.

\begin{figure}
  \centering
  \includegraphics[angle=90,width=0.8\columnwidth]{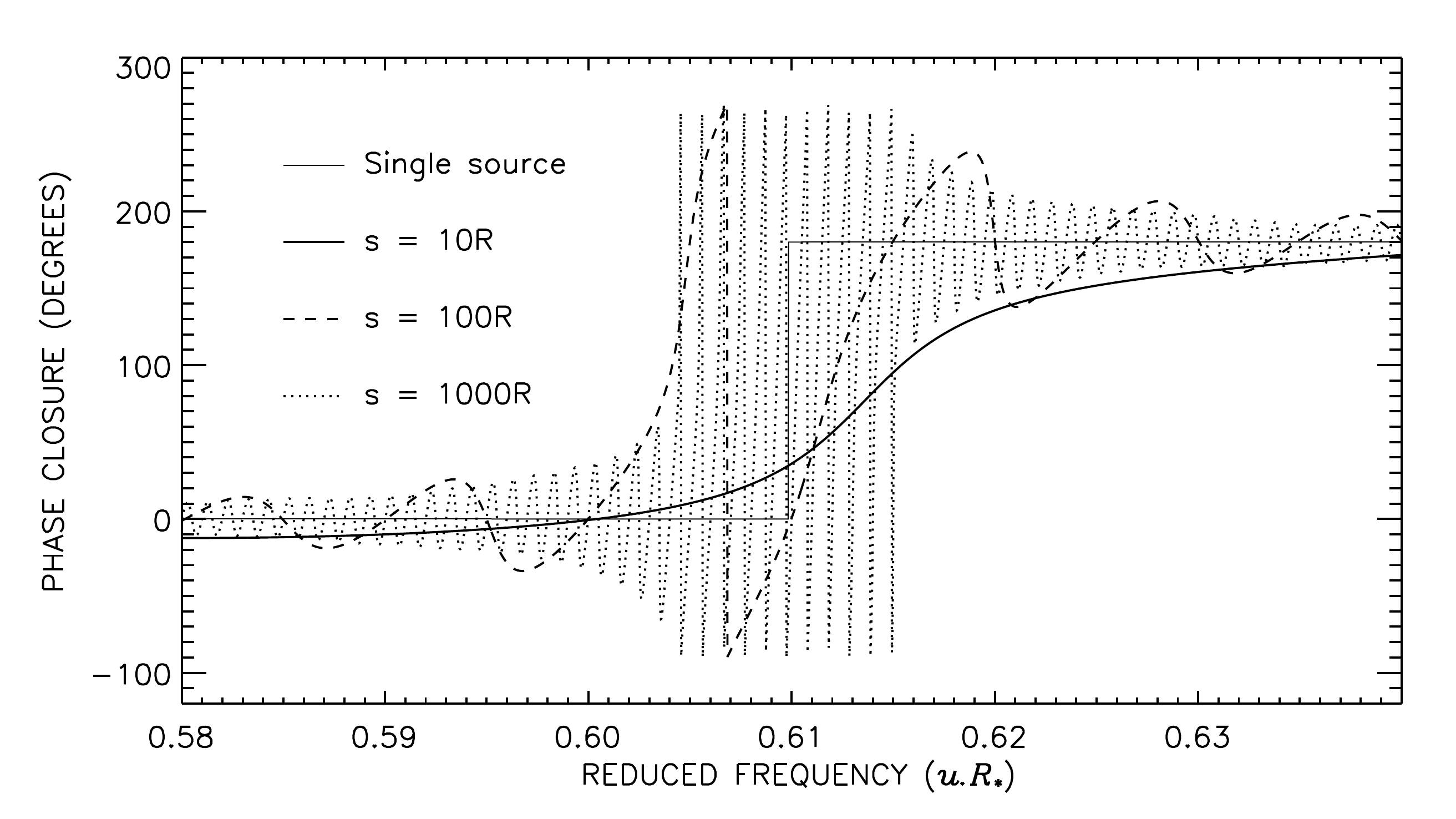}
  \caption{Closure phase modulo $2\pi$ of a double system, formed by
    an extended uniform disk and a point source, as a function of the
    maximum reduced frequency. The spectral resolution is
    $\mathcal{R}=1500$. The 3 frequencies \{$u_{23}, u_{12},
    u_{13}$\} (see Eq.~\ref{bispectrum}) have been chosen in the
    ratio 1,2,3, with the maximum frequency around the first zero of
    visibility of the primary star. The closure phase is displayed
    for 3 separations ($10\,R_\star$, $100\,R_\star$, and
    $1000\,R_\star$) and a flux ratio of $0.01$. The signature of the
    secondary source is dominant in the region around the minimum
    visibility of the primary (near reduced frequency $0.61$). The
    thin line with a 180 degree shift at 0.61 corresponds to the
    closure phase of the primary alone.}
  \label{signature}
\end{figure}
Fig.~\ref{signature} shows the closure phase around the first
visibility null of the primary, from a double system formed by an
extended uniform disk and an unresolved companion with a flux ratio of
1\% at various separations. Note the importance of the closure phase
signature from the companion. PCN is all about observing closure
phases near such nulls, where the correlated flux of the star is
vanishing, and fitting in the closure phase shape a stellar disk and
companion, that is, three basic parameters, $R_*$, $r$ and $s$. We
note that, as shown in Fig.~\ref{d0h0k0}, in various observational
cases with two short baselines and a long one, the closure phase
equals the phase of the long baseline since the phases on the two short
baselines can be approximated by that of an unresolved stellar disk, i.e., zero.

\subsection{Performances}

The performances of PCN have been analyzed in Chelli et al.\
(2009)\cite{2009A&A...498..321C}. We summarize here the main
properties:
\begin{itemize}
\item $\phi_c$ being proportional to $r/V_{*}$, there is no gain in
  SNR in the null. One still have to integrate enough to get the
  signal out of photon and detector noises.
\item Fibered interferometers have a field of view of one Airy
  disk. The field of view combined to the necessity to resolve the
  parent star imposes a maximum recoverable separation of
  $s_{max}/R_*\approx\,B/D$, which translates to a maximum of
  $\sim100\,R_*$ for VLTI. \item In the photon noise regime, the error
  on the flux ratio is given by $\epsilon(r)\approx 3 / S\sqrt{K}$
  ($S$ being the Strehl ratio and $K$ the total number of photons),
  which is 3 times worse than direct detection.
\end{itemize}

\subsection{Proof of concept: PCN on $\sigma$\,Puppis}

\begin{figure}
  \centering
  \includegraphics[width=0.7\columnwidth]{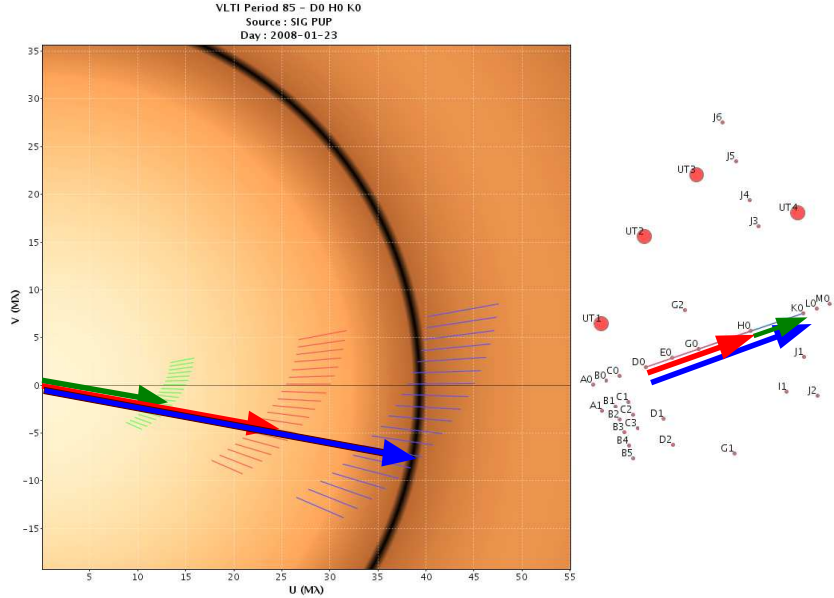}
  \caption{Sketch of the setup used at VLTI for PCN observations. The
    right panel shows the stations of the VLTI and the D0-H0-K0
    triplet used. The left panel shows the $(u,v)$ plane and a
    2-dimension map of the amplitude of the Fourier transform of a
    star of $6.4$\,mas ($\sigma$\,Puppis). The green, red and blue
    lines correspond to the baselines H0-K0, D0-H0 and D0-K0
    respectively, the latter sweeping above the first null (black
    circle) of the star due to the combination of the supersynthesis
    effect and the spatial resolution spanned by the observational
    bandwidth.}
   \label{d0h0k0}
  \end{figure}

  Although sampling quite sparsely the transition region around the
  first null, the $\sigma$ Puppis observations depicted in
  Figs.~\ref{figure1} and \ref{d0h0k0} have been interpreted in terms
  of PCN by Duvert et al.\ (2010)\cite{2010A&A...509A..66D}. The
  author best fit was $R_*=3.23\pm0.015$\,mas, a magnitude difference
  of $5.3\pm0.2$ between the giant and its companion, and a projected
  distance at time of observation of $~11$\,mas, that is, 4 stellar
  radii. With these numbers, assuming that the companion is a
  main-sequence star, they were able to estimate in two independent
  ways the masses in the system, both estimate giving the same result:
  a $5M_\odot$ K5III primary and a A2V $2.2M_\odot$ secondary.

\section{Science Cases}

There are a number of science cases that can benefit from the
possibility to detect, to very high dynamics, the presence of a faint
companion near a bright star, i.e., inside the airy disk patch of
light that a star produces in a conventional telescope:
\begin{itemize}
\item PCN can be complementary of spectroscopic diagnostics of
  binarity, in the cases when the inclination or period of the double
  system become unfavourable\cite{2010A&A...512A..39M} for an example
  of limitations of radial velocity detection of planetary
  companions).
\item PCN can perform the spectroscopy of the companion, when the S/N
  per channel is sufficient, the companion spectrum being the stellar
  spectrum multiplied by the flux ratio r estimated for each spectral
  channel.
\item For already know spectroscopic binaries of type SB1 with a giant
  component that can be spatially resolved by the interferometer, a
  single PCN detection can give the nature of the companion, and
  subsequently the mass of the giant star. The details of the method
  used are to be found in Duvert et
  al. (2010)\cite{2010A&A...509A..66D}.
\item PCN can help in finding Brown dwarves companions, and in general
  extend binarity statistics towards the low mass ratio limit.
\item finally, in favourable cases, one could expect to detect Hot
  Jupiters and at least measure their orbit and inclination.
\end{itemize}
A rough estimate for the number of targets feasible with the VLTI are
of $\approx100$ stars in $K$ and $\approx500$ in J.

\section{Observational Campaign 2009}

\begin{figure}[t]
  \centering
  \includegraphics[width=0.5\hsize]{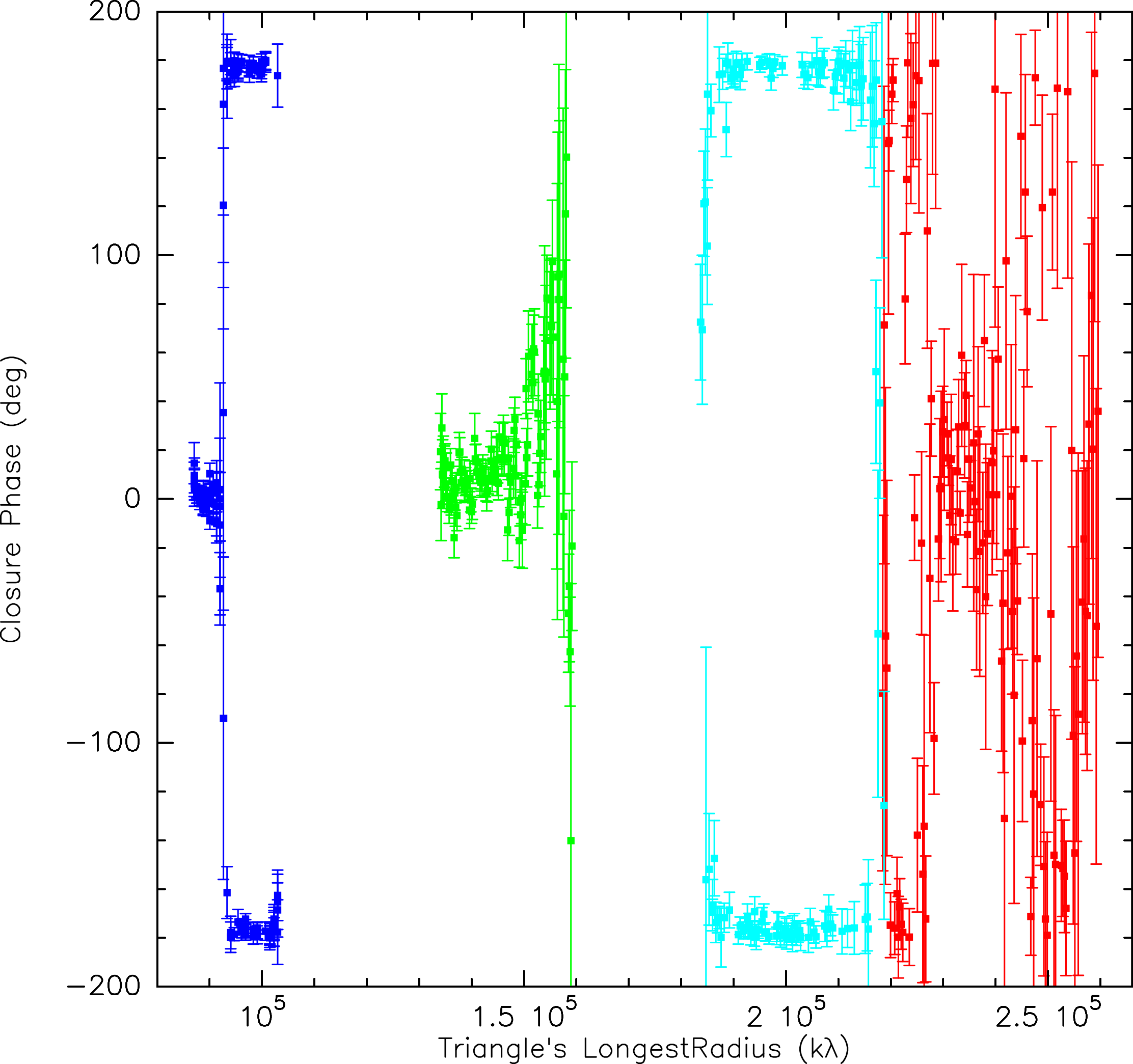}
  \caption{CHARA/MIRC observation of the star $\iota$\,Dra in July 2009, with
    several crossings of the null with closure phase jumps. The plot
    groups the phase closures on the 4 triplets in all the 24 channels
    of MIRC/GRISM (H band), as a function of the spatial frequency of
    the longest baseline in the triplet. From left to right: in blue, the
    triplet E1-E2-W2; in green, the E1-E2-W1; in light blue, E2-W1-W2;
    and in red, E1-W1-W2. }
  \medskip
  \label{fig:iotadra}
\end{figure}
We started an observational campaign in 2009 on two interferometric facilities:
AMBER/VLTI\cite{2007A&A...464....1P} with 3\,ATs, in
Medium-Resolution $K$ ($R=1500$), for which we got 11 nights, and, 
MIRC/CHARA\cite{2006SPIE.6268E..55M}, in $K$, $R=400$, for which
  we got 1 useful observing night (see Fig.~\ref{fig:iotadra}).

In the rest of the section, we focus on the AMBER/VLTI data. The
targets were of two types: SB1 stars with a giant primary, and, giant
stars for which no companion is known, which could serve either as a
mean to test the sensitivity limit, or eventually, would actually
possess a faint companion.

\subsection{Observational procedure}

The observational procedure was very simple.  We got an estimate of
the star diameter using the
\texttt{SearchCal}\cite{2006A&A...456..789B} web service. Given the
conservative uncertainties on the stellar diameters obtained this way,
the time of crossing of the first null was not very precisely
known. So we tried to stay as long as possible on the science star,
typically 4 to 5 hours, bracketing the observations with a 1\,h
observation of two calibrators, usually smaller, thus fainter objects,
giving much noisier measurements. Normally for closure phase, a
calibration is not really necessary. However in the case of AMBER, the
closure phase obtained on a star will contain the phase closure of the
so-called P2VM, the calibration matrix used to estimate the coherent
fluxes. It is thus necessary to correct this closure phase from the
one of a calibrator.

For the data reduction we used the new
\texttt{amdlib3}\cite{2009A&A...502..705C} algorithm which provides
accurate complex coherent fluxes.

We did not use the VLTI fringe tracker since it operates at $H$ band,
and, as can be seen in Fig.~\ref{d0h0k0}, if in $K$ the stars are fully
resolved at the distance of the longest baselines, in $H$ they are
already resolved at the distance of the second longest baseline, and
the small coherent flux at that point prevent a reliable fringe
correction: fringe trackers should operate at a longer
wavelength than the science observation!

\begin{figure}
  \centering
  \includegraphics[width=0.7\columnwidth]{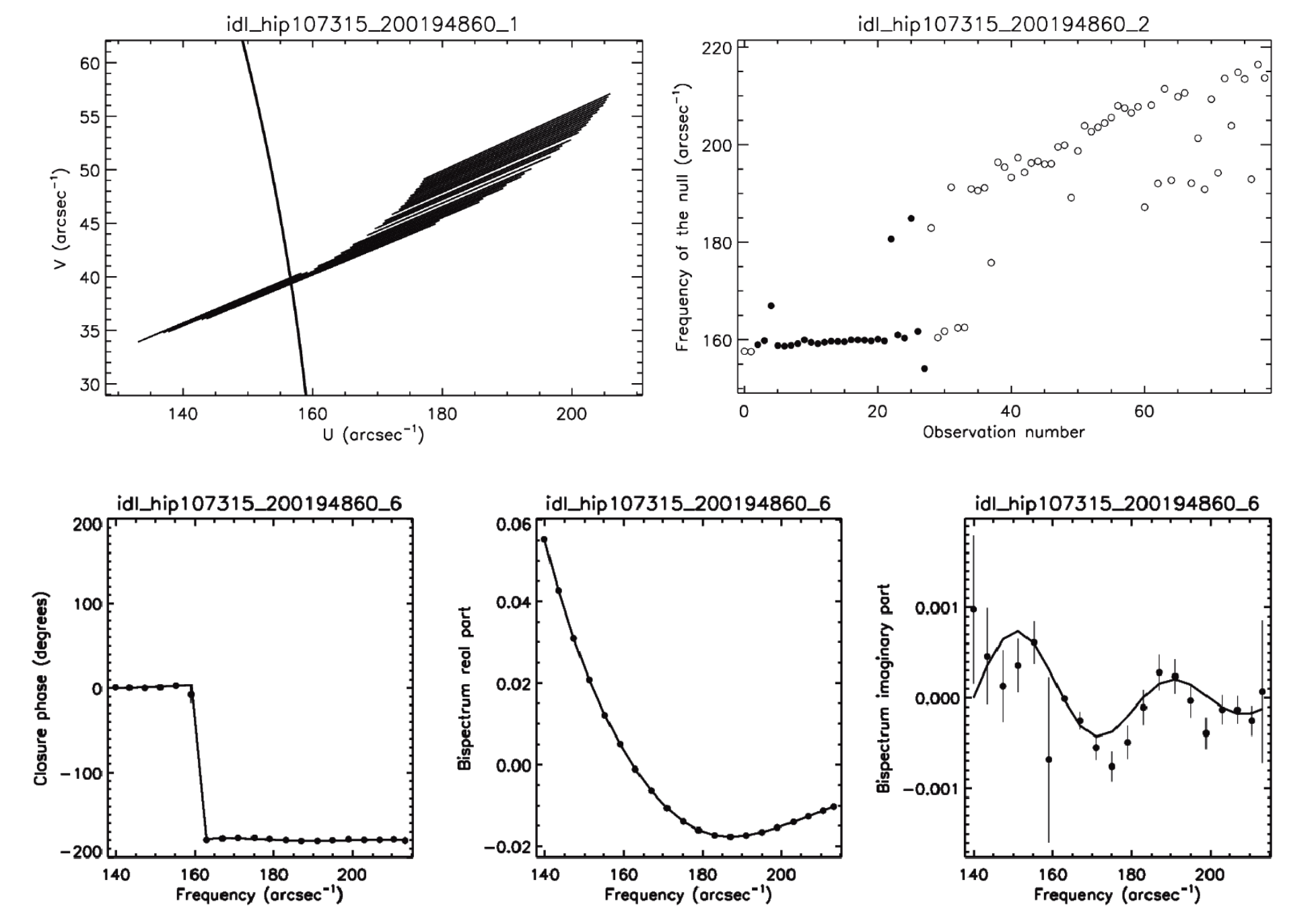}
  \caption{First results of PCN on HIP\,107315. See text for details.}
  \label{pcn_cal_fig}
\end{figure}

\subsection{Preliminary observations}

We focused firstly on the AMBER/VLTI data since we had previous experience of
such data with our $\sigma$ Puppis observations. Regrettably, AMBER observations
in 2009 suffer from an important closure phase noise, typically
$5^{\circ}$~rms. This comes from both an integration time per frame
($\approx100$\,ms) that is quite long with respect to the coherence time at $K$,
and from a piston-phase dependency induced by a dichroic in the VLTI
beams, optical element that has been replaced in November 2009. So,
for this campaign, we cannot achieve the ultimate in sensitivity
possible with AMBER/VLTI, and were obliged to develop new techniques
to reduce the data.

We have at the moment run only a general analysis on most of the
single stars in our observation list. This analysis consists in a
two-pass method. We first find the position of the zero of the real
part of the bispectrum, to get a first good estimate of the stellar
diameter. Then, we use this as an initial value to fit $R_*$,$s$ and
$r$. Figure~\ref{pcn_cal_fig} show the two main aspects of the process
for one of our objects, HIP\,107315. In panel (a), we show the $(u,v)$
coverage of the observations with the approximate position of the
null. One sees that in this case our observations cut the null at a
constant angle. The right (b) panel show the result of a the fit of
$R_*$, which gives extremely precise values for the star diameter in
all observations where the null is crossed (filled circles). From
these observations we derive $R_*=3.81875\pm0.018$\,mas. In the lower
three panels, we plot, from left to right, the shape of the closure
phase (c), the real part of the bispectrum (d) and the imaginary part
(e), where a sine-like signal of very small amplitude can be seen. If
this signal comes indeed from a companion (physically linked or field
star), the separation is not well constrained yet by the observations,
but the flux ratio can be reliably estimated to be
$3.3\,10^{-3}$. Thus, with the 2009 observations that suffer from a
large closure phase error, we can achieve already detections of faint
companions.

\subsection{Update on the AMBER closure phase stability}

\begin{figure}[t]
  \centering
  \includegraphics[angle=90,width=0.48\columnwidth]{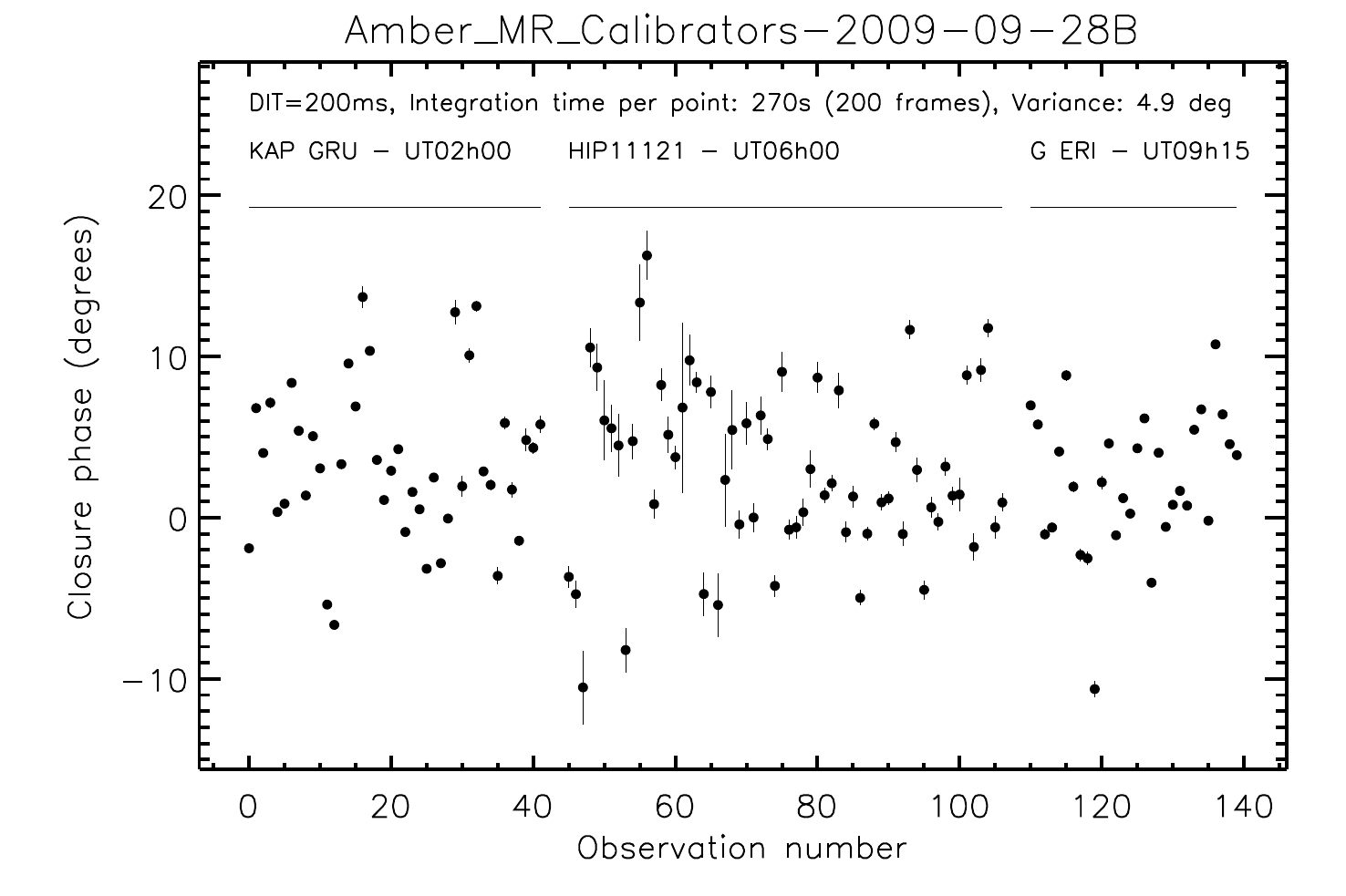}
  \hfill
  \includegraphics[angle=90,width=0.48\columnwidth]{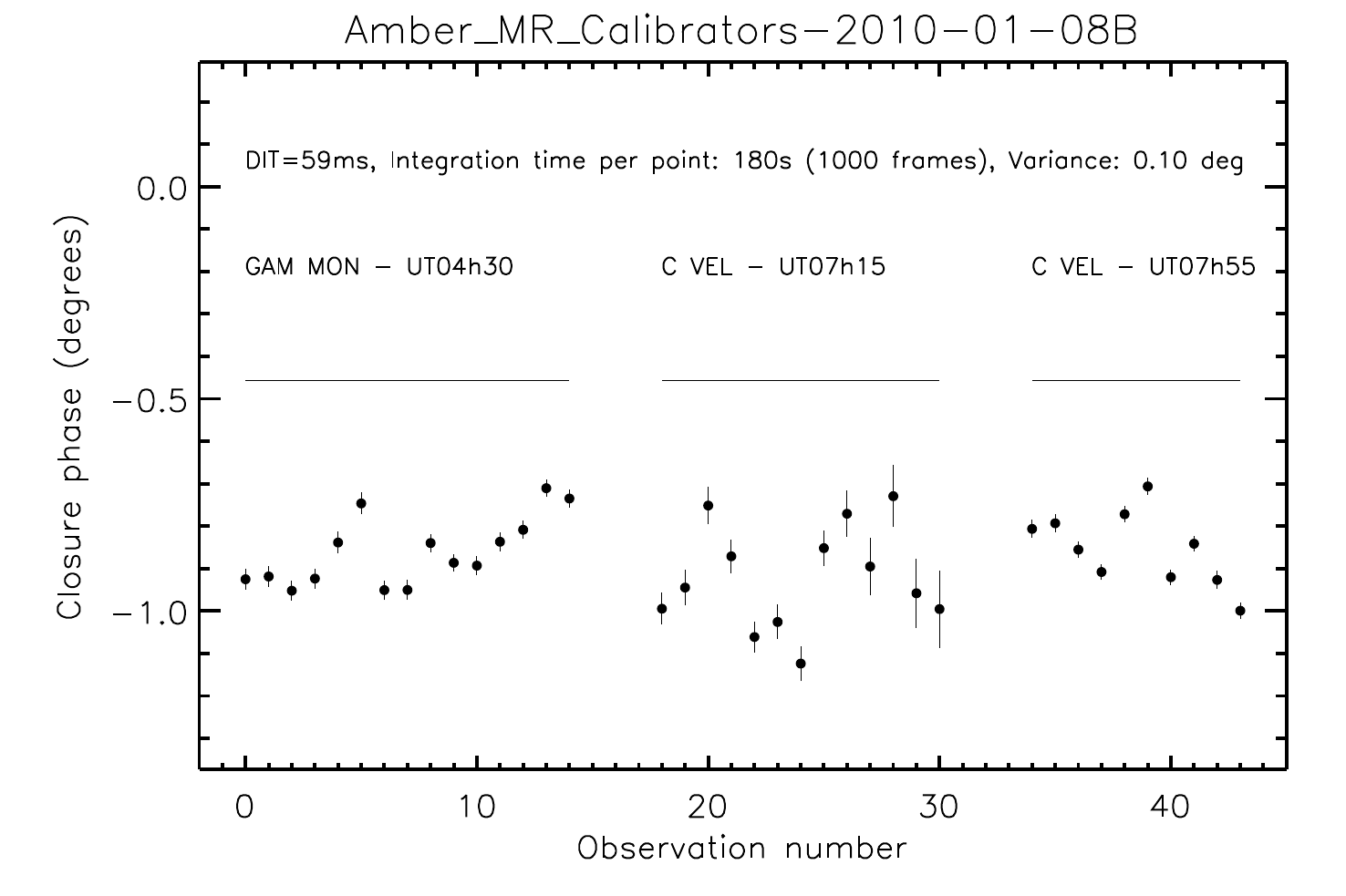}
  \caption{Improved stability of the closure phase of AMBER from 2009
    to 2010. Plots of the closure phase as a function of time for
    calibrators of different fluxes and at very different elevation in
    2009 (left) and in 2010 (right) in respectively 7\,h and 3\,h
    spans. The stability of AMBER has improved by a factor 50 from a
    variance of $4.9^{\circ}$ in 2009 to a variance of $0.1^{\circ}$
    in 2010. }
  \label{timestability}
\end{figure}
\begin{figure}[t]
  \centering
  \includegraphics[angle=90,width=0.48\columnwidth]{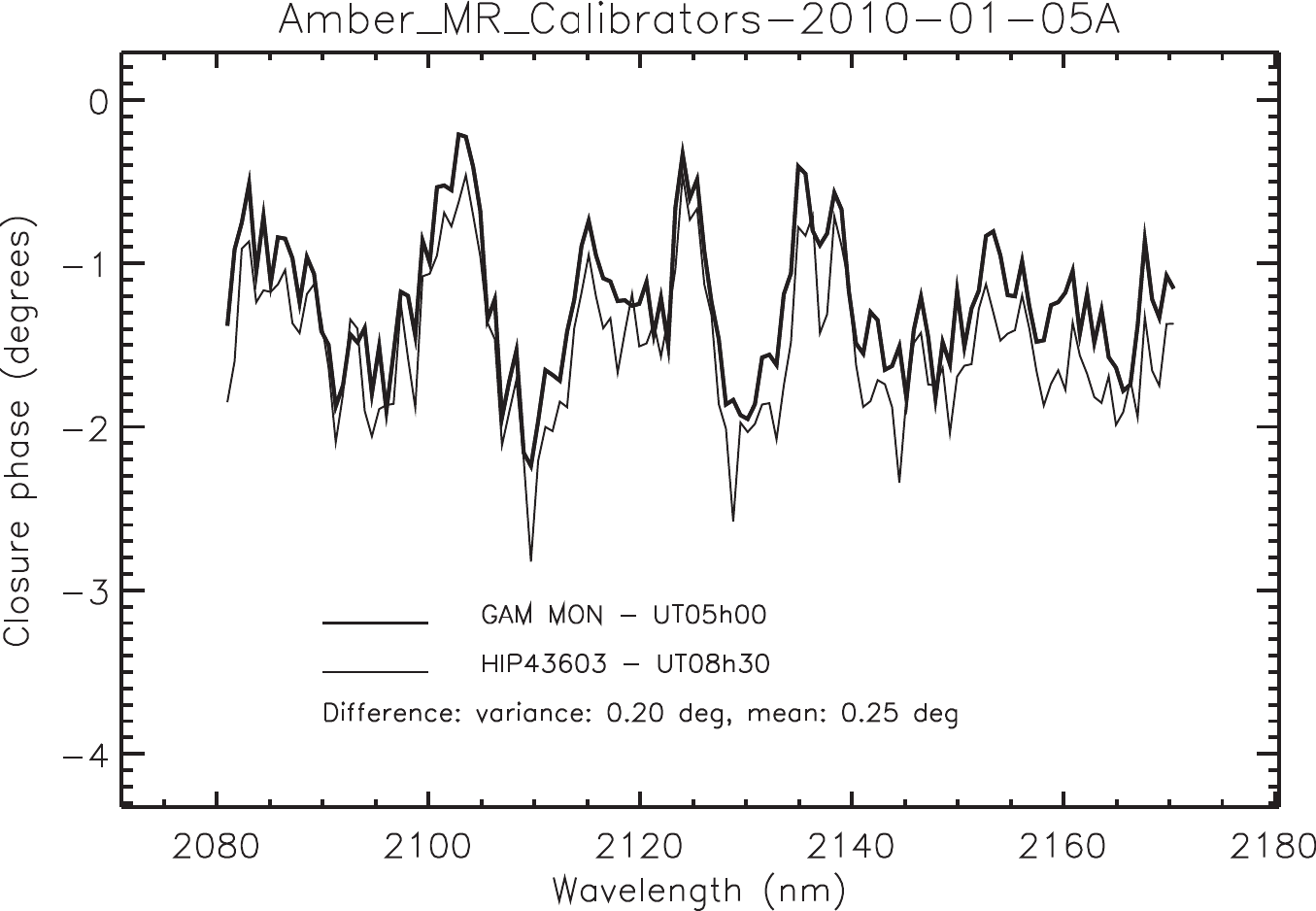}
  \caption{Improved stability of the closure phase of AMBER in 2010:
    plot of the closure phase as a function of wavelength for 2
    calibrators, of different fluxes and at very different elevation,
    taken at 4 hours intervals. The variance is $0.2^{\circ}$.}
  \label{spectrestability}
\end{figure}

The stability of the closure phase of AMBER has improved dramatically
since the replacement of the IRIS dichroic in the last months of
2009. We made new closure phase observations in early 2010, and, as
shown in Fig.~\ref{timestability}, the closure phase does not vary
from more than $0.1^{\circ}$ in a few hours (instead of $5^{\circ}$ in our 2009
observations). This is also visible in the spectral direction, as seen
in Fig.~\ref{spectrestability}, where the fixed phase artifacts are
stable and time and can be calibrated down to $0.2^{\circ}$ at least.

Although not fully understood, the closure phase dependence on the
piston induced by the previous IRIS dichroic was certainly related to
differential polarization effects. In this respect, it is remarkable
that our discovery of the PCN effect arose from the sole
unpolarized observations made with AMBER.

\section{Conclusion and Perspectives}

Phase Closure Nulling is an observational technique consisting in
measuring closure phases of resolved objects with a baseline triplet
where one of the baselines crosses a null of the object visibility
function. In these regions the signature of fainter scene objects are
seen in the closure phase, and stay out since the correlated flux of
the star is essentially nulled at these positions. PCN is limited in
field of view and needs quite high spectral resolution, but can
achieve detection of high-contrast objects. We report a first estimate
of contrasts down to $10^{-3}$ obtained in 2009 on AMBER with a bad
closure phase rms of $\approx5^{\circ}$. We think that the detection
limit can be pushed much farther thanks to the the excellent closure
phase stability ($\sim0.1^{\circ}$) displayed by AMBER since a
hardware change in VLTI that took place in November 2009.
 
\acknowledgments 
The authors warmly thank the ESO/VLTI and CHARA teams for their
support during the observations.


\bibliography{pcn}   

\begin{thebibliography}{1}

\bibitem{2009A&A...498..321C}
{Chelli}, A., {Duvert}, G., {Malbet}, F., and {Kern}, P., ``{Phase closure
  nulling. Application to the spectroscopy of faint companions},'' {\em
  \aap}~{\bf 498},  321--327 (2009).

\bibitem{VLT-TRE-AMB-15830-7120}
{Malbet}, F., {Duvert}, G., {Kern}, P., and {Chelli}, A., ``{February 2008
  AMBER Task Force run report},'' tech. rep., ESO Doc No.
  VLT-TRE-AMB-15830-7120, issue 1.2, dated 16/04/2008 (arXiv: 0808.1315)
  (2008).

\bibitem{1921ApJ....53..249M}
{Michelson}, A.~A. and {Pease}, F.~G., ``{Measurement of the diameter of alpha
  Orionis with the interferometer.},'' {\em \apj}~{\bf 53},  249--259 (1921).

\bibitem{2010A&A...512A..39M}
{Meunier}, N., {Desort}, M., and {Lagrange}, A., ``{Using the Sun to estimate
  Earth-like planets detection capabilities . II. Impact of plages},'' {\em
  \aap}~{\bf 512},  A39+ (2010).

\bibitem{2010A&A...509A..66D}
{Duvert}, G., {Chelli}, A., {Malbet}, F., and {Kern}, P., ``{Phase closure
  nulling of HD 59717 with AMBER/VLTI . Detection of the close faint
  companion},'' {\em \aap}~{\bf 509},  A66+ (2010).

\bibitem{2007A&A...464....1P}
{Petrov}, R.~G., {Malbet}, F., {Weigelt}, G., and {coll.}, ``{AMBER, the
  near-infrared spectro-interferometric three-telescope VLTI instrument},''
  {\em \aap}~{\bf 464},  1--12 (2007).

\bibitem{2006SPIE.6268E..55M}
{Monnier}, J.~D., {Pedretti}, E., {Thureau}, N., {Berger}, J.-P.,
  {Millan-Gabet}, R., {ten Brummelaar}, T., {McAlister}, H., {Sturmann}, J.,
  {Sturmann}, L., {Muirhead}, P., {Tannirkulam}, A., {Webster}, S., and {Zhao},
  M., ``{Michigan Infrared Combiner (MIRC): commissioning results at the CHARA
  Array},'' in [{\em Society of Photo-Optical Instrumentation Engineers (SPIE)
  Conference Series}{\nolinebreak\hspace{0.1em}]},  {\em Presented at the
  Society of Photo-Optical Instrumentation Engineers (SPIE) Conference} {\bf
  6268} (2006).

\bibitem{2006A&A...456..789B}
{Bonneau}, D., {Clausse}, J., {Delfosse}, X., {Mourard}, D., {Cetre}, S.,
  {Chelli}, A., {Cruzal{\`e}bes}, P., {Duvert}, G., and {Zins}, G.,
  ``{SearchCal: a virtual observatory tool for searching calibrators in optical
  long baseline interferometry. I. The bright object case},'' {\em \aap}~{\bf
  456},  789--789 (2006).

\bibitem{2009A&A...502..705C}
{Chelli}, A., {Utrera}, O.~H., and {Duvert}, G., ``{Optimised data reduction
  for the AMBER/VLTI instrument},'' {\em \aap}~{\bf 502},  705--709 (2009).

\end{thebibliography}
\bibliographystyle{spiebib}   

\end{document}